\documentclass[prb,a4paper]{revtex4}

\newcommand{\nc}{\newcommand}
\nc{\be}[1]{\begin{equation}\mbox{$\label{#1}$}}
\nc{\bea}[1]{\begin{eqnarray} \mbox{$\label{#1}$}}
\nc{\Section}[2]{\section{#2}\label{#1}}
\nc{\Bibitem}[1]{\bibitem{#1}}
\nc{\Label}[1]{\label{#1}}

\nc{\eea}{\end{eqnarray}}
\nc{\ee}{\end{equation}}

\nc{\bdm}{\begin{displaymath}}
\nc{\edm}{\end{displaymath}}
\nc{\dpsty}{\displaystyle}
\nc{\bc}{\begin{center}}
\nc{\ec}{\end{center}}
\nc{\ba}{\begin{array}}
\nc{\ea}{\end{array}}
\nc{\bab}{\begin{abstract}}
\nc{\eab}{\end{abstract}}
\nc{\btab}{\begin{tabular}}
\nc{\etab}{\end{tabular}}
\nc{\bit}{\begin{itemize}}
\nc{\eit}{\end{itemize}}
\nc{\ben}{\begin{enumerate}}
\nc{\een}{\end{enumerate}}
\nc{\bfig}{\begin{figure}}
\nc{\efig}{\end{figure}}

\nc{\arreq}{&\!=\!&}
\nc{\arrmi}{&\!-\!&}
\nc{\arrpl}{&\!+\!&}
\nc{\arrap}{&\!\!\!\approx\!\!\!&}
\nc{\non}{\nonumber}
\nc{\align}{\!\!\!\!\!\!\!\!&&}

\def\lsim{\; \raise0.3ex\hbox{$<$\kern-0.75em
      \raise-1.1ex\hbox{$\sim$}}\; }
\def\gsim{\; \raise0.3ex\hbox{$>$\kern-0.75em
      \raise-1.1ex\hbox{$\sim$}}\; }

\nc{\DOT}{\hspace{-0.08in}{\bf .}\hspace{0.1in}}
\nc{\Laada}{\hbox {$\sqcap$ \kern -1em $\sqcup$}}
\nc\loota{{\scriptstyle\sqcap\kern-0.55em\hbox{$\scriptstyle\sqcup$}}}
\nc\Loota{{\sqcap\kern-0.65em\hbox{$\sqcup$}}}
\nc\laada{\Loota}
\nc{\qed}{\hskip 3em \hbox{\BOX} \vskip 2ex}

\nc{\real}{{\rm I \! R}}
\nc{\Z}{{\sf Z \!\!\! Z}}
\nc{\complex}{{\rm C\!\!\! {\sf I}\,\,}}
\def\bigid{\leavevmode\hbox{\small1\kern-3.8pt\normalsize1}}
\def\id{\leavevmode\hbox{\small1\kern-3.3pt\normalsize1}}
\nc{\slask}{\!\!\!/}
\nc{\bis}{{\prime\prime}}
\nc{\pa}{\partial}
\nc{\na}{\nabla}
\nc{\ra}{\rangle}
\nc{\la}{\langle}
\nc{\goto}{\rightarrow}
\nc{\swap}{\leftrightarrow}

\nc{\EE}[1]{ \mbox{$\cdot10^{#1}$} }
\nc{\abs}[1]{\left|#1\right|}
\nc{\at}[2]{\left.#1\right|_{#2}}
\nc{\norm}[1]{\|#1\|}
\nc{\abscut}[2]{\Abs{#1}_{\scriptscriptstyle#2}}
\nc{\vek}[1]{{\rm\bf #1}}
\nc{\integral}[2]{\int\limits_{#1}^{#2}}
\nc{\inv}[1]{\frac{1}{#1}}
\nc{\dd}[2]{{{\partial #1}\over{\partial #2}}}
\nc{\ddd}[2]{{{{\partial}^2 #1}\over{\partial {#2}^2}}}
\nc{\dddd}[3]{{{{\partial}^2 #1}\over
    {\partial #2 \partial #3}}}
\nc{\dder}[2]{{{d #1}\over{d #2}}}
\nc{\ddder}[2]{{{d^2 #1}\over{d {#2}^2}}}
\nc{\dddder}[3]{{d^2 #1}\over
    {d #2 d #3}}
\nc{\dx}[1]{d\,^{#1}x}
\nc{\dy}[1]{d\,^{#1}y}
\nc{\dz}[1]{d\,^{#1}z}
\nc{\dl}[1]{\frac{d\,^{#1}l}{(2\pi)^{#1}}}
\nc{\dk}[1]{\frac{d\,^{#1}k}{(2\pi)^{#1}}}
\nc{\dq}[1]{\frac{d\,^{#1}q}{(2\pi)^{#1}}}

\nc{\bfT}{{\bf T }}

\nc{\cA}{{\cal A}}
\nc{\cB}{{\cal B}}
\nc{\cD}{{\cal D}}
\nc{\cE}{{\cal E}}
\nc{\cG}{{\cal G}}
\nc{\cH}{{\cal H}}
\nc{\cL}{{\cal L}}
\nc{\cO}{{\cal O}}
\nc{\cT}{{\cal T}}
\nc{\cN}{{\cal N}}
\nc{\cR}{{\cal R}}
\nc{\rvac}[1]{|{\cal O}#1\rangle}
\nc{\lvac}[1]{\langle{\cal O}#1|}
\nc{\rvacb}[1]{|{\cal O}_\beta #1\rangle}
\nc{\lvacb}[1]{\langle{\cal O}_\beta #1 |}
\nc{\bb}{\bar{\beta}}
\nc{\bt}{\tilde{\beta}}
\nc{\ctH}{\tilde{\cal H}}
\nc{\chH}{\hat{\cal H}}
\nc{\1}{\aa}
\nc{\2}{\"{a}}
\nc{\3}{\"{o}}
\nc{\4}{\AA}
\nc{\5}{\"{A}}
\nc{\6}{\"{O}}
\nc{\al}{\alpha}
\nc{\g}{\gamma}
\nc{\Del}{\Delta}
\nc{\e}{\textrm{e}}
\nc{\eps}{\epsilon}
\nc{\lam}{\lambda}
\nc{\Om}{\Omega}
\nc{\ve}{\varepsilon}
\nc{\mn}{{\mu\nu}}
\nc{\vp}{\varphi}

\nc{\rf}[1]{(\ref{#1})}
\nc{\nn}{\nonumber \\*}
\nc{\bfB}{\bf{B}}
\nc{\bfv}{\bf{v}}
\nc{\bfx}{\bf{x}}
\nc{\bfy}{\bf{y}}
\nc{\vx}{\vec{x}}
\nc{\vy}{\vec{y}}
\nc{\oB}{\overline{B}}
\nc{\oI}{\overline{I}}
\nc{\oR}{\overline{R}}
\nc{\rar}{\rightarrow}
\nc{\ti}{\times}
\nc{\slsh}{\hskip-5pt/}
\nc{\sm}{Standard~Model~}
\nc{\MP}{M_{\rm Pl}}
\nc{\mpl}{M_{\rm Pl}}
\nc{\tp}{t_{\rm Pl}}

\nc{\pmin}{p_{\rm min}}
\nc{\pmax}{p_{\rm max}}
\nc{\fo}{f_0}
\nc{\foi}{f_{0,i}\,}
\nc{\fop}{f_0^P}
\nc{\fou}{f_0^U}

\nc{\eff}{{\rm eff}}
\nc{\MT}{M_{\rm T}}
\nc{\ML}{M_{\rm L}}
\nc{\kk}{\vek{k}}
\nc{\pp}{{\rm p}}
\nc{\pt}{\partial_t}
\nc{\half}{{1\over 2}}
\nc{\w}{\omega}
\nc{\uhat}{\hat{U}_\w}

\nc{\etal}{\mbox{\it et al.}}
\nc{\ie}{{\it i.e. }}
\nc{\eg}{{\it e.g. }}
\nc{\trh}{T_{\rm RH}}
\nc{\ad}{{a'\over a}}
\nc{\bd}{{b'\over b}}
\nc{\Rd}{{R'\over R}}
\nc{\diag}{{\textrm{diag}}}
\nc{\mato}[1]{\tilde{#1}}
\nc{\sech}{\textrm{sech}}
\nc{\I}{\textrm{I}}
\nc{\II}{\textrm{II}}
\nc{\III}{\textrm{III}}
\nc{\vev}[1]{\langle #1 \rangle}
\nc{\hyp}{\,\; F_{1{\hskip -16pt}2}{\hskip 11pt}}
\nc{\brhom}{\overline{\rho}_M}
\nc{\brho}{\overline{\rho}}
\nc{\rhob}{\overline{\rho}}
\nc{\Pb}{\overline{P}}
\nc{\bH}{\overline{H}}
\nc{\ep}{{1+4\eps}}

\nc{\lcdm}{$\Lambda$CDM}
\nc{\ms}{\langle\sigma\rangle}

\def\smiley{\hbox{\large$\bigcirc$\hspace{-.80em}%
\raise.2ex\hbox{$\cdot\cdot$}\kern-.61em    
\lower.2ex\hbox{\scriptsize$\smile$}}\ }

\def\frowney{\hbox{\large$\bigcirc$\hspace{-.80em}%
\raise.2ex\hbox{$\cdot\cdot$}\kern-.635em
\lower.2ex\hbox{\scriptsize$\frown$}}\ }

\begin{document}

\title{\bf Inhomegeneous cosmological models and fine-tuning of the initial state}

\author{Peter~Sundell}
\author{Iiro~Vilja}
\affiliation{Department of Physics and Astronomy, University of Turku, 
FI-20014 Turku, Finland}
\date{\today}

\begin{abstract}
Inhomogeneous cosmological models are often reported to suffer from a fine-tuning problem because of the observer's location. We study if this is a generic feature in the Lema\^{i}tre-Tolman (LT) models, by investigating if there are models with freedom in the initial state. In these cases, the present fine-tuned location would be evolved from a non-fine-tuned initial state and thus vanishing the problem. In this paper, we show that this is not a generic problem and we give the condition when the LT models do not have fine-tuned initial state. The physical meaning of this condition, however, requires more investigation.  We investigate if this condition can be found from  a special case:  homogeneous models with matter, dark, and curvature density as parameters. We found that with any reasonable density values, these models do not satisfy this condition and thus do not have freedom in the initial state. We interpret this to be linked with the fine-tuning problem of the initial state of the homogeneous models, when the early time inflation is not included to them. We discuss of the condition in the context of non-homogeneous models.
\end{abstract}

\maketitle

\section{Introduction}

Supernovae Ia (SNIa) observations \cite{Riess1998, Perlmutter1999} made just 
before the break of the millennium implies that the universe appears to be expanding 
at an increasing rate. Observations of cosmic microwave background 
(CMB) radiation \cite{Spergel} supports the conclusions made from the supernovae 
observations. The most popular way to explain these observations are with models based on
Friedmann-Lema\^{i}tre-Robertson-Walker (FLRW) metric, which is based on general 
relativity and the cosmological and the Copernican principles. The cosmological 
principle merely states that the universe is spatially homogeneous everywhere, 
whereas the Copernican principle states that there are no preferred points in the 
universe. In the literature of cosmology, the  FLRW  metric and models based on it are extensively presented \cite{KolbTurner1994,Weinberg2008}.

Even though the FLRW-based models fit well inside the frame provided by our observations, 
they are   known to have problems, for example, the fine-tuning
problem \cite{KolbTurner1994, Weinberg2008} and the cosmological constant problem 
\cite{KolbTurner1994, Weinberg2000}. One of the strengths of the FLRW-based models 
is  simplicity due to various approximations, although, that can also be counted as a
weakness. It is also questionable if all approximations are made acceptably, as is 
pointed out by Shirokov and Fisher \cite{ShirokovFisher1963}, who questioned 
if the homogeneity approximation should be done to the Einstein tensor 
$G_{\mu \nu}$, rather than to the metric $g_{\mu \nu}$, since in general 
$<G_{\mu \nu}(g_{\mu \nu})>\neq G_{\mu \nu}(<g_{\mu \nu}>)$. However, this problem is 
more complex. As pointed out by Shirokov and Fisher 
\cite{ShirokovFisher1963}, Einstein equations are no longer tensor equations after 
averaging in a sense, that they can not be changed {\it e.g.} from covariant form to 
contravariant form with metric tensor without altering the equations. From this 
perspective it seems that only tensors rank 0 and scalars have well-validated 
averages, on which Buchert's approach is based\cite{Buchert1999}.
It is very possible that the problems  FLRW-based models have are due to 
approximations, and the work of Shirokov, Fisher and Buchert implies that 
the first approximation to question is the homogeneity. Such a model was first 
introduced by Lema\^{i}tre \cite{Lemaitre}, and later studied by Tolman \cite{Tolman} 
and it is called the Lema\^{i}tre-Tolman (LT) model. The model is further developed\cite{Bondi, BKHC2010}, but
it is not studied as widely as the  FLRW-based models and therefore all the 
problems it suffers have not been discovered.  Nevertheless, the LT model has already shown 
its power by overcoming an important problem in the FLRW-based models; the LT model can explain  SNIa observations without 
dark energy\cite{Garcia-BellidoHaugbolle2008,Mattsson,BlomqvistMortsell2010, Krasinski2013}.

In contrast to these promising results, also set backs to the LT models have been reported where the models are inconsistent with the observations when more observational constraints are imposed on  them \cite{MarraNotari2011, dePutterVerdeJimenez2012, Zumalacarregui2012, ZhangStebbins2011, MossZibin2011, BullCliftonFerreira2012, BiswasNotariValkenburg2010}. In these studies, however, void and bang time profiles (or their equivalents) are imposed on the system, decreasing the generality of the results. In addition, the resent tendency of the studies of the LT models has been such that  Earth is placed in the center of the spherical symmetry of the models. In \cite{AlnesAmarzquioui2006, BiswasNotariValkenburg2010, BlomqvistMortsell2010} were found  that the CMB observations give a better fit when we are located in the very vicinity of the center of the spherical symmetry.  As specific void and bang time profiles were used in these studies, thus making the results not general, a   fine-tuning issue caused by the observer's location is apparently created.

General constraints for the LT models are such where no forms for any functions, such as void or bang time, are assumed. Such constraints can be found by ruling out undesirable features of the system or assuming something desirable for the system.  A good example of a general  constraint ruling out an  undesirable feature is the  condition to avoid shell crossings \cite{HellabyLake1985}, since shell crossings are surfaces where the dust density diverges. In this article, we suggest a general condition for preventing fine-tuning issues in the inhomogeneous models. We discuss  this condition as generally as possible whenever the idea holds in all inhomogeneous models, but we derive the relevant equations in the LT framework. This condition arises by investigating when the structure of the equations governing the evolution 
of the universe is such, that it has an inherent property to make everything appear as observed, {\it i.e.},  the inhomogeneities of the universe at some point reach a state where they stabilize in a sense that the redshift observed from them no longer changes.

 Finding stability in any system usually relaxes at least some requirements of its initial state, motivating for the search of stability. When observing any system changing, by using a model, one can  calculate when and how it started. However, if one observes some system that shows no change, one can not say how and when it began.  Therefore, in the latter case  one can say  the initial conditions being relaxed  in a sense that from a large range of initial states, the system could have formed to the state it is  now observed. This allows freedom in the initial values of the system, therefore it is crucial  to demand that the state where the system shows no more change has been reached. Moreover, because we observe redshift, it makes most sense to study when the redshift no longer changes as the time passes.

 The assumption that the universe has reached a state where the redshift no longer changes can be reasonably satisfied by the filament-like structure of the large scale structure around us. As looking further in distance also means reflecting back in time, we see that the same kind of large scale structure have existed at least up to $ \sim 0.2$ in redshift\cite{Collessetal2001}. This does not mean that universe has reached some sort of stable state, but merely motivates to investigate the possibility. Even though we investigate redshift and this reasoning relies on the structure formation, which is governed by a  different set of equations, it is not unreasonable to expect them having some correlation, thus giving some validity to the deduction. 

Consider matter  inhomogeneities and their redshifts determined in a small solid angle. Assume a set of specific cases of the LT model that fit the observations within observational inaccuracies. Tracking back in time of the solutions of these specific cases leads us to the initial conditions from which the system could have developed.  If we also assume the stability discussed above has been reached, one can imagine at least, it relaxing the time of  birth of the universe at different locations, giving us a set of specific cases where the bang time is not fixed and can vary from place to place. On the other hand, looking into a small solid angle in some other direction, we will find matter inhomogeneities distributed somewhat differently, but the difference in the observed redshift is negligible. If we find a set of specific cases in the first direction, there should be no reason not to find another set of specific cases to fit these observations, and further on, another set of initial values.  This can be done in all directions, which suggests that almost any combination of proper over- and under dense regions (or shells in the case of the LT model) should produce the observed redshift-distance relation. This implies that the density function at the bang time is not very restricted either. Naturally, this  train of thought suggests that the density function is very restricted on the  surface of the last scattering due to the isotropicity of the CMB. However, if the birth time of the universe is relaxed at location where the surface of the last scattering takes place, it also enables the initial density distribution to be relaxed there, because different time periods allow different density distributions to develop into the same state. As a consequence, in general,  different directions in the sky can be described with different void and bang time profiles. Unfortunatelly this cannot be modelled with spherically symmetric models, but the apparent fine-tuning problem caused by the observers location is   due to the limitations of the spherical symmetric models. This is the main idea in this work.

In addition to avoiding fine-tuning issues, our approach offers another desirable feature to the system. 
In the  FLRW-based models, the observations indicate the universe to be very flat, creating a fine-tuning problem, which can be eliminated with early times inflation. This procedure cannot be generalized to the LT models directly, because 
the evolution of the universe in the LT models is dependent at coordinate distance, which would make 
the inflation occur differently in separate locations and the consequences of this are unknown. Of course, one can require  the LT model to approach the homogeneous limit in the early times\cite{BlomqvistMortsell2010, dePutterVerdeJimenez2012}, in which case the thoroughly studied FLRW  results can be utilized there, \emph{e.g.},   homogeneous inflation can be expected to take place. Nevertheless, demanding the universe to approach the homogeneous limit excludes many possible models.  In our approach, the system can not suffer from this problem by its construction. Hence, at least for this reason, inflation is not required in our approach.

Indeed, it is in itself interesting enough to study inhomogeneities with redshifts unaltered by time, even if the reason for this is unknown. The possible specific models arising with inhomogeneous bang time and not fine-tuned initial density function, are in themselves interesting enough to study these phenomena. After all, there might also be other effects, besides the one given earlier, causing the redshift not to change in time and thus relaxing the parameter values of initial state. These other effects, if they exist, can be expected to be found when specific models are studied carefully.

We begin  in Section II by introducing the  LT model from its most relevant parts to this paper. In Section III is shown in detail how the redshift changes at different distances in observer time in the LT models. The main result of this paper is given in Sec. IV, where the general requirement for the LT models to have a non-fine-tuned initial state is formulated. We go to the homogeneous limit in Section V and show how the relevant equations reduce in this case. In Section VI  we show how the methods constructed can be utilized by applying them to the realistic homogeneous cases. The paper is concluded and discussed in Section VII.

\section{The Lema\^{i}tre-Tolman model}

The LT model\footnote{We consider the special case of the model where the observer is at the origin.} describes a dust filled inhomegeneous 
but isotropic universe, which energy momentum tensor reads as 
$T^{\mu \nu}=\rho u^{\mu} u^{\nu}$. Here $\rho=\rho(t,r)$ 
is matter density and $u^{\mu}$ is the local four-velocity. As the coordinates are 
assumed to be comoving, the four-velocity is simply $u^{\mu}=\delta^{\mu}_t$. 
The standard synchronous gauge metric in the LT model is given 
by \footnote{We use units in which the speed of light $c=1$.}
\be {LT_metric}
ds^2=-dt^2+\frac{R_r^2 dr^2}{1+f}+R^2 \left( d \theta^2 +
\sin^2 \theta\, d\phi^2 \right),
\ee
where the subscript $r$ denotes partial derivative with respect to the radial coordinate, 
$R_r=\partial R(t,r)/\partial r$, $R=R(t,r)$, and $f=f(r)> -1$. In this prescription, the 
evolution of universe is built in to the local scale factor $R$, whereas function $f$
controls the overall radial dilatation. Including the cosmological constant $\Lambda$
into the Einstein equations, after some integrations, one obtains the relevant
differential equations as 
\be {GFE}
R_t^2=\frac{2M}{R}+f+\frac{\Lambda}{3}R^2,
\ee 
and 
\be {DEQ}
\kappa \rho=\frac{2M_r}{R_r\,R^2},
\ee
where $R_t=\partial R(t,r)/\partial t$, $M=M(r)$ is an arbitrary function, $\kappa=8\pi G/c^4$, and $G$ is the 
Newton's constant of gravity. 

The conventional definition of redshift is
\be{redshift}
1+z=\frac{c \delta_0}{c \delta_e},
\ee
where $\delta_0$ and $\delta_e$ are the observed and emitted oscillation time periods of the light wave, respectively. In the case of the LT model, the following expression can be derived  from the redshift definition: \cite{MustaphaHellabyEllis1998,MustaphaBassettHellabyEllis1998,Mattsson} 
\be  {dz/dr}
\frac{dz}{dr}=\frac{(1+z)R_{rt}(t(r),r)}{\sqrt{1+f}},
\ee 
where $t(r)$ describes how a light ray propagates in space, {\it i.e.}, the path of a radial light ray given by 
the radial null geodesic, where $ds^2=d\theta^2=d\psi^2=0$. Using metric (\ref{LT_metric}), it is described by the differential equation
\be {null_geodesic}
\frac{dt}{dr}=\pm \frac{R_r }{\sqrt{1+f}},
\ee
where the signs correspond to an incoming ($-$) and   outgoing ($+$) light rays.

\section{The Change of redshift}

From the redshift equation (\ref{dz/dr}) can the redshift on the null geodesic for a given function $t(r)$  be solved uniquely. This is best represented from the integral form of Eq. (\ref{dz/dr}):
\be  {rs3}
\int_0^{z_e} \frac{d\bar{z}}{1+\bar{z}}=\int_{0}^{r_e} \frac{R_{rt}(t(r),r)}{\sqrt{1+f(r)}}dr=-\int_{t(0)}^{t(r_e)} \frac{R_{rt}(t,r(t))}{R_{r}(t,r(t))}dt,
\ee
where $r_e$ is the comoving distance of the observable, $t(r_e)$ is the time of emission,  $t(0)$ is the time of observation and $z_e$ is the redshift corresponding to these parameters. The last equality in  (\ref{rs3})  follows from Eq. (\ref{null_geodesic}).  From (\ref{rs3}), we see that, in general, different functions $t=t(r)$ corresponds to different emission and observation times for a fixed $r_e$, but always can be found two functions $t(r)$ and $\tilde{t}(r)$ for which $t(r_e)=\tilde{t}(r_e)$ and $t(0)=\tilde{t}(0)$. On the other hand, if $t(r_e)=\tilde{t}(r_e)$ for all $r_e \in [0,\infty )$, functions are equivalent to each other and  describes the same physical situation. 

The differential equation  (\ref{dz/dr})  describes how the redshift changes with respect to the coordinate distance at given time determined by the function $t(r)$, thus by solving (\ref{dz/dr}) with an indefinite integral over both sides would yield a functional form of $z(r,t(r))$ for the redshift. Let us next see how the redshift of objects  at some fixed distance $r_e$  changes as it is observed at different times. Two light crests are emitted at the times $t(r_e)$ and  $t(r_e)+\delta_e$ and observed at the times $t(0)$ and  $t(0)+\delta_0$ respectively, describing two different redshifts of the same object received at different times, 
\be {ze}
\left\{ \begin{array}{l l} 
z_e&:=z(r_e,t(r_e),t(0)) , \\ 
\tilde{z}_e&:=z(r_e,t(r_e)+\delta_e,t(0)+\delta_0) .
\end{array}  \right.
\ee
 The difference between these redshift values divided by the difference of the observing times, $\delta_0$, gives an evaluation  how the redshift of the  observable at $r_e$ changes next. Eq. (\ref{dz/dr}) implies that the redshifts $z_e$ and $\tilde{z}_e$ are obtained by choosing functions  $t(r)$ and $\tilde{t}(r)$ appropriately:  $t(r)$ and $\tilde{t}(r)$ need to have correct observation and emission times corresponding to the radial coordinate distance\footnote{In general, we could choose functions  $t$ and $\tilde{t}$ to be dependent on differently scaled coordinate distances, {\it i.e.}  $t(r)$ and $\tilde{t}(\tilde{r})$, so that  $r \neq \tilde{r}$. However, the coordinate degree of freedom allows us choose  $r= \tilde{r}$, which we apply here for simplicity.}.   Eqs. (\ref{redshift}) and (\ref{dz/dr}) show that by choosing
\be {tildet}
\tilde{t}(r)=t(r)+\delta(t(r),r),
\ee
where
\be {deltar}
\delta(t(r),r)=\frac{\delta_0}{1+z(t(r),r)},
\ee
 we find the desired redshifts $z_e$ and $\tilde{z}_e$. For the system to be coherent, we should have
$\tilde{t}(r_e)=t(r_e)+\delta(t(r_e),r_e)$ for all $r_e \in [0,\infty)$. Based on our earlier discussion, this requirement uniquely determines functions $t(r)$ and  $\tilde{t}(r)$  and hence  $\delta(t(r),r)$ too.

The change of redshift is thus
\be {deltaz}
\delta z_e  := \tilde{z}_e-z_e =\exp\left[ \int_{0}^{r_e} \frac{R_{rt}(\tilde{t}(r),r)}{\sqrt{1+f(r)}}dr  \right]-\exp\left[ \int_{0}^{r_e} \frac{R_{rt}(t(r),r)}{\sqrt{1+f(r)}}dr  \right] 
 \approx (1+z_e)\int_{0}^{r_e} \frac{R_{rtt}(t(r),r)}{\sqrt{1+f(r)}(1+z) }dr \delta_0 ,
\ee
where the last step follows from the first order approximation with respect to $\delta_0$ after the Taylor expansion of $R_{rt}(\tilde{t}(r),r)$. Integration over the redshift in the integrand can be executed after the redshift is presented with respect to $r$, which can be obtained by solving the differential equation (\ref{dz/dr}) by integrating over intervals $[0,z]$ and $[0,r]$.

We have arrived in a novel  way to the same result published in \cite{YooKaiNakao2011}. Another difference from their paper is that we began from a more general situation by starting from functions $t(r)$ and $\tilde{t}(r)$ and showing that in this particular physical situation, $\tilde{t}(r)$ is uniquely determined as $\tilde{t}(r)=t(r)+\delta(t(r),r)$. Eq. (\ref{deltaz}) can also be achieved by yet another way, which is by giving a similar treatment as above to the time dependent form of the Eq. (\ref{rs3}). This is a longer way compared with the one above, but important because that it implies the system to be coherent and well understood.

The change of redshift in observer time, often referred as the redshift drift, can be used effectively in two distinct ways. Eq. (\ref{deltaz}) can be used so that $r_e$ is the only variable, in which case (\ref{deltaz}) gives an evaluation how the redshift is going to  change at all distances on the null geodesic next. This can be used to discriminate models from each other in the future, after the accuracy of observations has improved enough\cite{YooKaiNakao2011}. The differential limit of Eq. (\ref{deltaz}) is derived by  dividing both sides of it with $\delta_0$  and going to the limit $\delta_0 \rightarrow 0$, yielding
\bea{YKN}
\frac{dz(r_e,t(r_e,t_0),t_0)}{dt_0}=[1+z(r_e,t(r_e,t_0),t_0)]\int_{ 0}^{r_e}\frac{R_{ttr}(t(r,t_0),r)}{\sqrt{1+f(r)} }\frac{1}{1+z(r,t(r,t_0),t_0)}  dr   ,
\eea
where $t(0)=t_0$, giving us a differential equation of $z_e(t_0)$, $r_e$ being a parameter. Since Eq. (\ref{YKN}) is interpreted as a  differential equation, the explicit dependence on the observation time  is also written; from the derivation of Eq.(\ref{deltaz}), one can see that only the redshift and $t$   depend on $t_0$,  and more specifically, the   dependence on $t_0$ in redshift comes  through $t$ and the lower integration limit. Note, that this also applies to the definition (\ref{ze}) and we have used here the notation $z_e=z(r_e,t(r_e,t_0),t_0)$.

\section{Converging redshift} \label{converging redshift}

When the redshift drift equation is considered as a (integro-)differential equation, it can also be treated as one. However, when considering it describing a dynamical system, one has to be cautious with deductions and interpretations as it is not describing the whole system but rather every distance  separately.   These issues are presented in the Conclusions and Discussions.

As discussed in the Introduction, we are interested in the situations where the  redshift of  observables at some given coordinate distance no longer changes in time, thus we investigate when the redshift  solved out  from the differential equation  (\ref{YKN}),  converges towards some constant value in observer time.
For  this, we construct an autonomous system of Eq. (\ref{YKN}) and then study the behavior of the system in phase plane\cite{Strogatz1994, Percival1982}.  Eq. (\ref{YKN}) is not autonomous as it explicitly depends on $t_0$, but by artificially considering $dz_e/dt_0$ to be implicitly dependent on $t_0$ through $z$ and identity function, an autonomous system of this physical situation is described by two first order differential equations: Eq. (\ref{YKN}) and the trivial relation $dt_0/dt_0=1$.
Clearly fixed points, where both $dz_e/dt_0$ and $dt_0/dt_0$ are  zero, can not be found. Hence, finite stable points do not exist in the autonomous system. However, it is possible for $dz_e/dt_0$ to converge towards zero as as the observer time goes to infinity. Since we do not expect $1+z_e$ to vanish, this can only take place when the integral of Eq. (\ref{YKN}) converges to zero. Thus, if
\be{req1}
 \lim_{t_0 \rightarrow \infty}\int_{ 0}^{r_e}\frac{R_{ttr}(t(r),r)}{\sqrt{1+f(r)} }\frac{1}{1+z(r,t_0)}  dr=0, 
\ee 
the redshift of the observables at distance $r_e$ converges towards some constant value. Qualitatively, there are two different cases
for this to take place: as the observer time goes to infinity, the integrand can either go to zero or change sign  appropriately allowing the integral over it to vanish. Both of these circumstances are controlled by the function $R_{ttr}$, since we do not expect $1+z$ or $\sqrt{1+f}$ to change sign or to approach  infinity over time.  
The former case occurs when
\be{req2}
\lim_{t_0 \rightarrow \infty} R_{ttr}(t(r),r)=0 \qquad \text{for all } r \in [0,r_e],
\ee
and thus holds for all distances up to $r_e$ ensuring the redshift drift to disappear for every object up to $r_e$. The latter case makes the drift disappear only at specific distances; consider  $R_{ttr}(t(r),r)$ crossing over the $r$-axis in   $(r,R_{ttr})$-plane  ones. There can be only one $r_e$ value for which the integral of the Eq. (\ref{YKN}) disappears. This generalizes:  if  $R_{ttr}(t(r),r)$ crosses over the $r$-axis $n$ times, then the maximum amount of $r_e$ values for which the integral of the Eq. (\ref{YKN}) dissapears is $n$.  According to this, the autonomous system can  have many observables with different distances where $dz/dt_0 $ converges to zero if $R_{ttr}$ is an oscillating function with respect to $r$. From (\ref{GFE}), we find  $R_{ttr}$  to take an elegant form even in the presence of the cosmological constant:
\be{Rttr}
R_{ttr}=-\frac{M_r}{R^2}+2\frac{M R_r}{R^3}+\frac{2 \Lambda}{3}R_r.
\ee

 In general it is expected to be very challenging  locating the distances $r_e$ which satisfies (\ref{req1}). Therefore, it is reasonable to begin by studying when $R_{ttr}=0$ on the null geodesic. This can be done using the present null geodesic, since
in this paper we assume that the observed redshifts are not coincidental, but the universe has already reached a state where these observed values have already reached the values they are converging towards. After determining when  $R_{ttr}=0$ on the present null geodesic we have a better understanding how many zero redshift drift locations there can be and at what coordinate range they are located. This facilitates finding the locations where the redshift drift is zero. After the zero drift loci are traced down, one needs to check if these loci stay as such when $t_0 \rightarrow \infty$.  In practice, this can be a very difficult task since the differential equations are solved numerically. For this reason, in this case, it is sufficient to study when the integral disappears for long enough time.

\section{The Homogeneous limit}

When going to the homogeneous limit, the form of the line element needs to be chosen to be compatible with the chosen gauge. For example, the standard  FLRW metric relation between $r$ and $z$ is\cite{KolbTurner1994}
\be {ih1}
r=\frac{2 z \Omega _m^{\text{now}}+(2 \Omega _m^{\text{now}}-4) \left(\sqrt{z \Omega _m^{\text{now}}+1}-1\right)}{a_{\text{now}} H_{\text{now}} 
(z+1)(\Omega
   _m^{\text{now}})^2},
\ee 
where $ \Omega _m^{\text{now}}$,  $a_{\text{now}}$ and $H_{\text{now}}$ are the present  matter density, scale factor and Hubble constant. However, if the gauge degree of freedom is used to give $t(r)=t_0-r$, then the relation between $r$ and $z$ is
\be {ih2}
r=\pm \sqrt{\frac{1}{k}-\frac{a_ {\text{now}}^2}{k(1+z)^2}}, 
\ee
which is obtained from  from Eq. (\ref{null_geodesic}) and   $a=a_0/(1+z)$ when they are evaluated at the present moment. 
Clearly Eqs. (\ref{ih1}) and (\ref{ih2}) are not equivalent, 
except in some special cases. We denote the present time (the time since the big bang) by $t_ {\text{now}} $ and observation time by $t_0$,   $t_ {\text{now}} $ being a constant and $t_0$ being a variable. Also, quantities with super- or subscript 'now' indicates that they are evaluated at the time $t_ {\text{now}} $ and thus are constants too. Similarly,  quantities with super- or subscript '0' means that they are evaluated at the time $t_ 0 $ and are also variables. In the special case where the observation is done today, the observation time has the value  $t_0\equiv t_ {\text{now}} $.

In spherically symmetric homogeneous space-time, 
the metric can always be given as\cite{Weinberg1972}
\be {eq4}
ds^2=g(v) dv^2+h(v)\left(d\textbf{u}^2+\frac{k(\textbf{u}\cdot d 
\textbf{u})^2}{1-k\textbf{u}^2} \right),
\ee
where $v$ and $\textbf{u}=(u_1, u_2, u_3)$ are the coordinates, $g(v)$ is a 
negative and $h(v)$ is a positive function of $v$, and $k$ is spatial curvature 
and can be chosen to be 1, 0, or -1. For our purposes, it is convenient to define the
new coordinates $t$, $r$, $\theta$, and $\varphi$ by
\bea {eq5}
 \sqrt{-g(v)}dv&=&dt, \nonumber\\
u_1&=&\Theta (r) \sin \theta \cos \varphi, \nonumber\\
u_2&=&\Theta (r) \sin \theta \sin \varphi, \nonumber\\
u_3&=&\Theta (r) \cos \theta. 
\eea
Then we have
\be {IH_metric}
ds^2=-dt^2+a^2\left[ \frac{\Theta_r^2}{1-k \Theta^2}dr^2+
\Theta^2 d\Omega^2 \right],
\ee 
where $a=a(t)=\sqrt{h(v)}$, $\Theta=\Theta (r)$, $\Theta_r=\partial \Theta(r)/\partial r$  and $d\Omega^2=d\theta^2+\sin^2\theta d\varphi^2$.
With the above metric definition equations (\ref{dz/dr}) and 
 (\ref{null_geodesic})  reduces to be
\bea{dz/dr_H}
\frac{dz}{dr}=\frac{(1+z)a_t \Theta_r}{\sqrt{1-k\Theta^2}} ,
\eea
where $a_t=\partial a(t)/\partial t$ and
\be {null_geodesic_H}
\frac{dt}{dr}=- \frac{a \Theta_r }{\sqrt{1-k\Theta^2}}.
\ee 
Combining these, the well-known relation in the homogeneous models, 
\be{dz/dr_HI2}
\frac{1}{1+z}=\frac{a}{a_0},
\ee
is reproduced. The  inconsistency presented  between Eqs.  (\ref{ih1}) and (\ref{ih2}) can thus be prevented by choosing the function $\Theta(r)$ appropriately. On the null geodesic $\Theta(r(z))$ can be integrated out from Eq. (\ref{null_geodesic_H}) using Eq. (\ref{dz/dr_HI2}) once $a(t)$ or $\dot{a}(a)$ is  explicitly known, giving for different  curvature cases:
\be{eq9}
\Theta(r(z_e)) = \left\{  \begin{array}{ll} \sinh \left( \int_{0}^{z_e} \frac{1+z}{a_0  z_t}dz \right), \quad 
\quad & k=-1\,, \\
 \int_{0}^{z_e} \frac{1+z}{a_0  z_t}dz, \qquad \qquad \qquad & k=0\,, \\
\sin \left( \int_{0}^{z_e} \frac{1+z}{a_0  z_t}dz \right), \quad & k=1\, ,
\end{array} \right. 
\ee
where  $\Theta(r=0)=0$ is determined by the requirement $R(z=0)=0$ and  $z_t=\partial z/\partial t$.

With the line element (\ref{IH_metric}), Eq. (\ref{GFE}) takes the normal Friedmannian form and
 therefore can be written as
\be {at}
a_t=\pm H_{\text{now}} a \sqrt{ \Omega_m^{\text{now}} (a/a_{\text{now}})^{-3}+ \Omega_{\Lambda}^{\text{now}} + \Omega_k^{\text{now}} (a/a_{\text{now}})^{-2}  },
\ee
where   $\Omega_{\Lambda}^{\text{now}}$  is the present value of the dark energy density and  $\Omega_k^{\text{now}}$ is the present value of the spatial curvature density, which can be expressed as $ \Omega_k^{\text{now}}=1- \Omega_m^{\text{now}} - \Omega_{\Lambda}^{\text{now}} $.

The LT redshift drift equation (\ref{YKN}) simplifies considerably in the homogeneous case. By using the null geodesic equation (\ref{null_geodesic})
and going to the homogeneous limit yields
\bea{YKNh}
\frac{dz}{dt_0}=-(1+z(r_e,t_0))\int_{t_0}^{t_e}\frac{a_{tt}(t)}{a(t) }\frac{1}{1+z(r(t),t_0)}  dt=   -\frac{a(t_0)}{a(t_e)}\int_{t_0}^{t_e}\frac{a_{tt}(t)}{a(t) }\frac{a(t)}{a(t_0)}  dt=   \frac{a_{t}(t_0)-a_{t}(t_e)}{a(t_e)}.
\eea
Similarly as in the general case, this equation can be used in two different ways: to evaluate how redshift is going to drift next on the null geodesic or as an ordinary differential equation. In the former case, $t_0$ is a parameter and $t_e$ is the only variable, whereas in the latter case both of them are considered as variables and are related via the null geodesic equation (\ref{null_geodesic_H}) \cite{Sandage1962}. This is more transparent by writing the null geodesic equation in integral form
\be{0drift2}
\int_{t_0}^{t_e}\frac{dt}{a(t)}=u,
\ee
where $u$ is a constant\footnote{Here $u$ is constant because we are studying only one object in comoving coordinates.} , which explicit form depends on the curvature:
\be{u}
u=\left\{ \begin{array}{l l} 
\text{arcsinh} \, \Theta, & \quad k=-1, \\ 
 \Theta, & \quad k=0, \\ 
\arcsin \Theta, & \quad k=1 .
\end{array} \right.
\ee

\section{example: the realistic homogeneous cases}

In this section, we demonstrate how the methods presented in this paper can be utilized in practice to find locations where the redshift drift is zero at the moment and if they stay as such as time elapses. Methods are applied to the limiting homogeneous cases in which the evolution is described by (\ref{at}) and there is no pressure in the system.

In the homogeneous universes $R_{ttr}=a_{tt}\Theta_r$ and the function $\Theta_r$ is related to the coordinate distances and thus can not be zero besides in the origin, hence it is sufficient to study when $a_{tt}$ vanish. In the absence of  pressure
\be{att}
a_{tt}(t)=-\frac{4\pi}{3}\rho(t) a(t)+\frac{\Lambda}{3} a(t),
\ee
where $\rho(t)=\rho_{\text{now}} [a_{\text{now}}/a(t)]^3$ and $\rho_{\text{now}}$ is the present matter density. Trivially, $a_{tt}$ is zero in empty space, but we ignore this possibility due to the fact that space containing only curvature cannot describe our universe. In non-empty space,  $a_{tt}$  can vanish on the today's observer's null geodesic only at the distances $r'>0$ where
\be{0drift}
\frac{ a(t_{\text{now}})}{ a(t(r'))}=\left(\frac{\Lambda}{4\pi \rho_{\text{now}}} \right)^{1/3}.
\ee
For a monotonic $a(t(r))$, this can take place only at one coordinate distance. In realistic cases, the function $a(t(r))$ can be assumed to be monotonic and we justify this at this section's end.
Hence, there can exist only one $r_e$, where the redshift drift disappears, and if so, it is found where the integral in (\ref{YKN}) disappears. In the homogeneous case, this reduces down to equation $a_t(t_0)=a_t(t_e)$ (see (\ref{YKNh})), which by using Eq. (\ref{at}) becomes
\bea{0drift0}
a_0\sqrt{\Omega_m^{\text{now}} \left( \frac{a_{\text{now}}}{a_0} \right)^3+\Omega_{\Lambda}^{\text{now}}+(1-\Omega_m^{\text{now}}-\Omega_{\Lambda}^{\text{now}})\left(\frac{a_{\text{now}}}{a_0}\right)^2}= \\
a_e\sqrt{\Omega_m^{\text{now}} \left( \frac{a_{\text{now}}}{a_e} \right)^3+\Omega_{\Lambda}^{\text{now}}+(1-\Omega_m^{\text{now}}-\Omega_{\Lambda}^{\text{now}})\left(\frac{a_{\text{now}}}{a_e}\right)^2}.
\eea
This equation reduces even further:
\be{0drift5}
a_0^2\Omega_{\Lambda}^{\text{now}}+\frac{a_{\text{now}}^3\Omega_{m}^{\text{now}}}{a_0}=a_e^2\Omega_{\Lambda}^{\text{now}}+\frac{a_{\text{now}}^3\Omega_{m}^{\text{now}}}{a_e}.
\ee

On today's observer's null geodesic, where $t_0=t_{\text{now}}$, we find three loci, $r_e$, where the redshift drift is zero: one trivial, one unphysical and
\be{0drift4}
 a(t(r_e))=
a_{\text{now}} \frac{- \Omega_{\Lambda}^{\text{now}} + \sqrt{\Omega_{\Lambda}^{\text{now}}} \sqrt{\Omega_{\Lambda}^{\text{now}} + 4  \Omega_{m}^{\text{now}}}}{2 \Omega_{\Lambda}^{\text{now}}}. \, \,   
\ee
For $\Omega_{\Lambda}^{\text{now}}=0.7$ and $\Omega_{m}^{\text{now}}=0.3$, the locus where the redshift drift is zero for today's observer is where $z \approx 2$.

However, the redshift drift does not stay zero at this locus for models where $\Omega_m^{\text{now}}$ and $ \Omega_{\Lambda}^{\text{now}} $ are non-negative and where $a(t(r))$ is a monotonic function.
 For the cases where either $\Omega_{\Lambda}$ or $\Omega_m$ is zero, this can directly be seen from  Eq. (\ref{0drift5}), since it yields $a(t(r_e))=a(t(0))$, which can only be satisfied at the origin for a monotonic $a(t(r))$. The case where neither $\Omega_{\Lambda}$ nor $\Omega_m$ is zero, we compare  Eqs. (\ref{0drift2}) and  (\ref{0drift5}).  The comparison between these equations is executed since the former describes how $t_0$ and $t_e$ are related  and the latter describes how they should be related for redshift drift to stay zero at specific loci as the observer time elapses. Since we can find three solutions for $a_e$ from Eq. (\ref{0drift5}), we should find the same solutions for $a_e$ from Eq. (\ref{0drift2}) too if the equations are equivalent. Furthermore, we should find the same derivatives $da_e/dt_e$ from both of the equations. After derivating both equations, with respect to $t_e$, and solving the terms  $da_e/dt_e$ from them, it is evident that the derivatives are not the same in general. However, by setting the derivatives equal, we find the constraint when they are equivalent. This yields the equation
\be{eq33c}
(2 a_0  \Omega_{\Lambda}^{\text{now}}- \Omega_{m}^{\text{now}}/a_0^2)/(2 a_e  \Omega_{\Lambda}^{\text{now}} -  \Omega_{m}^{\text{now}}/a_e^2) = a_e/a_0,
\ee
from which  $ \Omega_{\Lambda}^{\text{now}}$ can be solved:
\be{eq34}
 \Omega_{\Lambda}^{\text{now}} = -\frac{ \Omega_{m}^{\text{now}}}{2 a_0 a_e (a_0 + a_e)}.
\ee
Evidently, the derivatives $da_e/dt_e$ of Eqs.  (\ref{0drift2}) and  (\ref{0drift5}) cannot be the same for positive matter and  dark energy density. Therefore, Eqs  (\ref{0drift2}) and  (\ref{0drift5}) do not have the same realistic solutions and hence no realistic homogeneous model has a location where the redshift drift stays zero.

 The monotonicity of $a(t(r))$ can be verified by investigating when
\be{eq33}
0=\frac{da(t(r))}{dr}=a_t(t(r)) t_r(r)=\pm  \frac{  H_{\text{now}}a^2 \Theta_r \sqrt{ \Omega_m^{\text{now}} (a/a_{\text{now}})^{-3}+ \Omega_{\Lambda}^{\text{now}} + \Omega_k^{\text{now}} (a/a_{\text{now}})^{-2}  } }{\sqrt{1-k\Theta^2}}  ,
\ee
where Eqs. (\ref{null_geodesic_H}) and (\ref{at}) are used. Clearly, the function inside the square root is positive if $| \Omega_{k}^{\text{now}}|<  \Omega_{\Lambda}^{\text{now}},  \Omega_{m}^{\text{now}}$ and  $a=0$ only at the Big Bang.  In addition, the gauge degree of freedom is still unused, so we can choose $\Theta=r$. Thus, $a(t(r))$ is zero only at the Big Bang time and hence is a monotonic function.

As was shown, at the homogeneous limit, we can find locations with zero redshift drift, but the redshift  in these loci is not converging to any constant  in non-empty space. Therefore, in the perspective of our analysis, in every location the observed redshift is coincidental as there are no positive distances where the change stays zero.

\section{Conclusions and discussion}

We consider when the  initial conditions of the inhomogeneous cosmological models can be relaxed without assuming additional features to the models, \emph{e.g.} inflation. We conclude that this can be done by studying when the change of redshift of objects converges towards zero while the observer time approaches to infinity. Analysis of the phase space $(z,z_{t_0})$ is used to find this out and the analysis is restricted to the LT models. The methods found here can be used to search for models with non-fine-tuned initial values and test models if their initial values can be relaxed. The latter is demonstrated on the present paper at the homogeneous limit and the former is planned to do for future studies.

The deduction that some  initial conditions can be relaxed if the redshift drift is zero is, in detail, given in the Introduction, but was mainly motivated by the observer's point of view. If one observes some system changing, by using a model one can also calculate when and how it  started. However, if one observes some system that shows no change, one can not say how and when it began. For this reason, it is crucial for us to assume that the universe has already reached the state where the observed redshifts no longer change in time, as we want freedom to the initial state. Therefore, it is sufficient to study the convergence of the redshift only at the loci where the redshift drift is zero already on today's observer's null geodesic. Loci, where the redshift drift is zero on today's observer's null geodesic and stays zero as observer time approaches to infinity, we refer to the non-incidental as the observed redshift value is fixed there for today's and future's observers. Other loci we call incidental, since the observed redshift value changes there with time.

In section \ref{converging redshift}, we showed  that the redshift can converge towards a constant value in two qualitatively distinct ways; it can converge at every coordinate distance at some range or at a specific coordinate distances. The previous situation is mathematically more sound, because in this case, the redshift converges in every point at this range enabling non-fine-tuned initial conditions at every point at this range. Physically, however, it is difficult to understand how this can take place since inhomogeneities are encountered. The latter case is, at the physical point of view, more understandable but mathematically not as well founded, since the initial conditions can be expected to have some freedom only at the points where the redshift drift is zero. In the perspective of our approach, the observations from the loci with non-zero redshift drift have fine-tuned initial states. However, one can possibly over come this problem by  physical reasoning. The fact is that the observations are not uniformly distributed in distance,\emph{e.g.}, we can not observe the redshift closely behind mass concentrations. Hence, if the observations are mainly  from distances where the redshift drift is zero, the problem appears to vanish, at least if the incidental observations fit inside inaccuracies.   Nevertheless, this is only speculation and we have planned to clarify these issues in the subsequent papers. The point here is, however, that both of these cases have appealing aspects for which reason they can not be excluded \emph{a prior}, but require investigation in detail.

As discussed earlier, it is a necessity for us to require converging redshift values on the present null geodesic, but the physical meaning of these loci remains  to be determined. For this reason, it is difficult to say how distant these loci should reach. For example, if the redshift converges towards constant values at every coordinate distance at some range, what is this range? In the perspective of our analysis, it is reasonable to assume the range  includes at least the vicinity of the Milky Way, but how far out should it reach?  Following the reasoning given in Introduction, these loci should be found at least as far as we see filament-like large scale structure. However, structure formation and redshift drift are controlled by different equations, so this reasoning can merely be used as a first guideline.

In the Introduction, we discussed why the initial dust density distribution should have some freedom. It was deduced to hold when our line of sight in different directions does not show isotropy in dust distribution, like in large scale structure. However, CMB shows strong isotropy, thus in this light, the density distribution freedom on the surface of the last scattering is small. In fact, somewhat similar behavior exists when studying   so-called ``Giant Void"  models, which is a class of inhomogeneous LT models of which specific cases was successfully fitted to SNIa and CMB data. SNaI  data does not seem to  restrict the observer too close to the center of the void in these models, which is compatible with our deduction that the dust density distribution is not very restricted as far as filament-like structure can be detected. Furthermore, the CMB appears to be very restricting observation in these models, placing the observer at the very vicinity of the center of the void.\cite{AlnesAmarzquioui2006,BlomqvistMortsell2010,BiswasNotariValkenburg2010} This is again compatible with our conclusions. On the other hand, the range, in which we choose non-incidental redshifts  on our null geodesic to cover, can play a significant role here. Indeed, if the locus of the surface of the last scattering is non-incidental, the bang time at that locus is not fixed. Moreover, Giant Voids are not needed to explain observations if the bang time profile is appropriate \cite{Krasinski2013}. This suggests that by employing the bang time freedom on the surface of the last scattering, introduced by its non-incidentaly, would bring freedom to the density distribution of this location. Hence, one can  find generalized Giant Void models where the void is not spherically symmetric, and furthermore, not even a void in every direction. Of course, these type of models can no longer be modelled with LT models due to their spherical symmetricity. Thus, within the framework of the LT models, the apparent fine-tuning problem caused by the observer's location is  due to the limitations of the spherical symmetricity of the models.

As an illustration how to utilize methods constructed in this paper, we analyzed the homogeneous limit of the LT model in detail, of which one special case describes the concordance \lcdm \,  model at late times.  We showed that realistic homogeneous models including curvature, dust and dark energy can have locations on the present observer's null geodesic with zero redshift drift, but the feature in these locations cannot stay permanent. This means that there are no loci in these models where the redshift converges towards some constant value. Thus, all observed values are incidental. This result, however,  is not 
surprising; as it is known, without inflation, the homogeneous models have a fine tuning problem of matter and energy density \cite{KolbTurner1994, Weinberg2008}, and in this case even including the redshift effects in the equations cannot fix this problem. 

An interesting physical issue arose when studying the homogeneous case, which can be expected to be encountered with more general models too. Earlier we discussed if the  inhomogeneous models with a handful of non-incidental loci can relax initial conditions. The reasoning used no longer works if there exists only one or two of these loci. In the homogeneous models, loci with zero redshift drift were found, but none of these were non-incidental. However, it is possible that there are inhomogeneous models where we find one or two non-incidental locations. The physical meaning of this is unknown but interesting, thus we will look into that  in detail if encountered.

Even though it is known that any LT  model cannot describe the entire history of the universe due to the absence of the pressure, it can describe the late times. Consequently, as we investigate a late time feature, it can be expected to hold in more general models - especially in those approaching the LT models at the late times.

The overall view our analysis casts on the inhomogeneous models is interesting. We have constructed an effective method of finding inhomogeneous LT models, which do not have fine-tuning issues, at least when it comes to the bang time and density distribution functions. The caveat is that the physical meaning of our requirement for non-incidentally is not  fully understood. It might very well be that one cannot construct a physically  appealing system with non-incidental observations. On the other hand, the idea cannot be rejected without further investigation.

The inhomogeneous models are the last attempt to explain the universe with    Einstein's general relativity and the  well-known energy forms, dust, radiation and curvature. 
These models seem to have a statistically unfavourably feature, the  fine-tuning of the observer's location. If this truely turns out to be a generic property, assuming that the future observations do not  violate the Copernican principle,  it would make these models statistically inviable. In this case, the well-established physics cannot describe the nature without this disability and something extra, such as inflation, is required to overcome this issue. 
Hence, it is important to study the fine-tuning issue of the inhomogeneous models carefully.

\acknowledgments

This study is partially (P. S.) supported by the Magnus Ehrnroothin S\2\2ti\3 foundation: grant number FY2012n49. Robert M. Badeau, Ph.D, of the Turku University Language Centre, checked the linguistic accuracy of this paper.


\end{document}